\documentclass{MITcsail}

\newcommand{\be}{\begin{equation}}
\newcommand{\ee}{\end{equation}}
\newcommand{\bal}{\begin{aligned}}
\newcommand{\eal}{\end{aligned}}
\newcommand{\scriptD}{\mathcal{D}}
\newcommand{\scriptR}{\mathcal{R}}

\newcommand{\scriptX}{\mathcal{X}}
\newcommand{\true}{\text{true}}

\usepackage{graphicx}
\usepackage{comment}
\DeclareMathOperator*{\argmin}{arg\,min}

\title{Multiple testing for signal-agnostic searches of new physics with machine learning
}

\author
{
Gaia Grosso~$^{1, 2, 3}$\footnote{\href{mailto:gaiag795@mit.edu}{gaiag795@mit.edu}} and 
Marco Letizia~$^{4, 5}$\footnote{\href{mailto:marco.letizia@edu.unige.it}{marco.letizia@edu.unige.it}}\\
\normalfont{\small $^{1}$NSF AI Institute for Artificial Intelligence and Fundamental Interactions}\\
\normalfont{\small $^{2}$MIT Laboratory for Nuclear Science, Cambridge, MA}\\
\normalfont{\small $^{3}$School of Engineering and Applied Sciences, Harvard University, Cambridge, MA}\\
\normalfont{\small $^{4}$MaLGa--DIBRIS, University of Genoa, Genoa, Italy} \\
\normalfont{\small $^{5}$INFN, Sezione di Genova, Genoa, Italy}\\
}

\begin{document}

\maketitle
\thispagestyle{firstpagestyle} 

\begin{abstract}
In this work, we address the question of how to enhance signal-agnostic searches by leveraging multiple testing strategies. Specifically, we consider hypothesis tests relying on machine learning, where model selection can introduce a bias towards specific families of new physics signals. We show that it is beneficial to combine different tests, characterised by distinct choices of hyperparameters, and that performances comparable to the best available test are generally achieved while providing a more uniform response to various types of anomalies. 
Focusing on the New Physics Learning Machine, a methodology to perform a signal-agnostic likelihood-ratio test, we explore a number of approaches to multiple testing, such as combining p-values and aggregating test statistics. 
\end{abstract}


\section{Introduction}\label{sec:intro}
After decades of experimental results that contributed to the development and confirmation of the Standard Model of particle physics (SM), we are in a phase in which no compelling theoretical prediction is guiding experimental searches. It is then important to develop model-independent analyses that are potentially sensitive to new physics effects not necessarily predicted by any specific Beyond the Standard Model (BSM) scenario. This is an extraordinary difficult task given the complexity of collider data and the fact that new physics can manifest itself as a deviation from the SM predictions in infinitely many ways. Moreover, these effects are expected to be extremely rare (poor signal-to-background ratio) and/or hidden (characterising uncommon observables).

Several proposals for partial model-independent analyses have been applied to experimental data. Early instances, such as those in \cite{D0:2000vuh,CDF:2007iou,Choudalakis:2011qn,ATLAS:2018zdn,CMS:2020zjg}, were based on simplifying assumptions about the way new physics effects could appear in the measurements and, as a consequence, they were limited to a selection of interesting final states. Crucially, these methodologies focused on theoretically motivated high-level features to reduce dimensionality and resorted to traditional statistical techniques.

Recently, machine learning has been leveraged to design flexible and multivariate data-driven tests, further enhancing signal-agnostic strategies. Various approaches have been proposed over the past few years (the reader can find an exhaustive review in \cite{Belis:2023mqs}), some of which have already been applied to experimental data (see \cite{CMS:2024lwn} and \cite{ATLAS:2020iwa}). Despite their potential, the adoption of these techniques introduces new challenges, particularly in understanding how model selection can impact sensitivity and bias the analysis towards specific signal hypotheses.

Here, we address this topic considering as a case study the model introduced in \cite{Letizia:2022xbe}, where classifiers based on efficient kernel methods \cite{meanti2020kernel} are used to design a multivariate and unbinned likelihood-ratio test in which the alternative hypothesis is derived from data. This idea (dubbed The New Physics Learning Machine, NPLM for brevity) was initially presented in \cite{DAgnolo:2018cun} using neural networks. 
The approach to hyperparameters selection proposed in \cite{Letizia:2022xbe}
is based on a mix of statistical and heuristic criteria that was shown to work well empirically on a number of benchmarks (see also \cite{Grosso:2023ltd,Grosso:2023scl}). 
However, it is not guaranteed that the resulting model has optimal performance or a uniform response over a wide range of possible deviations from the reference expectation. It is then natural to ask whether it is possible to explore more principled and controlled approaches to model selection with the purpose of improving sensitivity as well as robustness.

In this work, we propose to improve model-independent searches by exploiting the idea of multiple testing \cite{Lehmann2022TestingSH,wasserman2013all}. Instead of selecting a single test based on a specific learning model, we propose to define multiple ones characterised by different choices of hyperparameters, implicitly defining different alternative hypotheses, and combine their outputs into a single meta-analysis while accounting for the look-elsewhere effect. We compare different methods and show that this framework results in a more uniform response over new physics effects of different nature, and that the achieved performance is comparable or close to the best model, which is not known a priori in real use cases.\footnote{The code to reproduce our results can be found in \url{https://github.com/mletizia/multiple-testing-nplm}.} This strategy has been proposed in recent studies \cite{schrab2023mmd,biggs2023mmdfuse}, in the context of a class of kernel-based two-sample tests known as Maximum Mean Discrepancy.

The paper is organised as follows. In Section~\ref{sec:2} we recall the underlying statistical framework and revise the NPLM approach to hypothesis testing in its implementation based on kernel methods, with a focus on the model selection pipeline. In Section~\ref{sec:3} we review the multiple testing problem, explore some approaches to address it and introduce how to integrate them in the NPLM methodology. Section~\ref{sec:4} is dedicated to numerical experiments. Concluding remarks are given in Section~\ref{sec:5}.

\section{The search for new physics as a signal-agnostic hypothesis test}\label{sec:2}
Let us consider a set $\scriptD=\{x_i\}_{i=1}^{N_\scriptD}$ of realisations of a random variable $x\in\scriptX\subseteq \mathbb{R}^d$ representing experimental measurements, independent and identically distributed according to an unknown true distribution $p_\true (x)$. We call \emph{reference distribution}, $p(x|R)$, the distribution of the events as predicted by a reference model R (in our case, the SM). Additionally, in high energy collider physics, the number of collected events $N_\scriptD$ is also a random variable, following a Poisson distribution characterised by a true expected value $N(\true)$. We name $N(R)$ the expected number of collected events as predicted by the reference model. Ideally, the analysis should also be sensitive to discrepancies between these two values. This can be formalised by introducing the quantity
\be\label{n_density}
n(x|\cdot)~=~N(\cdot) p(x|\cdot),
\ee
namely the probability density normalised to the associated expected number of events for any given physical theory.  
The goal of a signal-agnostic test is to determine whether $p(x|R)$ is a good description of the data without introducing alternative models. In statistical terms, this can be framed as a \emph{goodness-of-fit} (GoF) test. However, the reference distribution is commonly not available in closed form. In this work, we consider the situation in which a \emph{reference sample} $\scriptR=\{x_i\}_{i=1}^{N_\scriptR}$ can be obtained by simulations or with measurements from a control region. 
The problem is then to assess the goodness of the population-level null hypothesis 
\be
H_0: n_\true (x)=n(x|R)
\ee
from finite data, by comparing $\scriptD$ with $\scriptR$. A task of this kind is commonly known in statistics as a \emph{two-sample test}. In this framework, the alternative hypothesis is simply the negation of the null
\be
H_1: n_\true (x)\neq n(x|R).
\ee
In order to have an accurate description of the reference distribution, we assume that $N_\scriptR\gg N_\scriptD$.

A two-sample hypothesis test requires to choose a test statistic, namely a function
\be
t: \scriptX^{N_{\scriptD}}\times \scriptX^{N_{\scriptR}}\rightarrow \mathbb{R}
\ee
that maps the measured data and the reference sample to a measure of their compatibility defined as a real number $t_{\rm obs}=t(\scriptR,\scriptD)$. 
To establish the statistical significance of the outcome of the test, a p-value is computed. This quantity is the probability, under the null hypothesis, of observing values that are at least as extreme as the measured ones
\be
p_{\rm obs}= P(t\geq t_{\rm obs}|H_0).
\ee
The observed p-value is then compared to a predefined threshold $\alpha\in [0,1]$, representing the highest acceptable rate of false positives associated with the test, i.e. the probability of rejecting the null hypothesis if true. A discovery is claimed if $p_{\rm obs}<\alpha$. These probabilities can be mapped to \emph{Z-scores} using the following expression
\be\label{eq:zscore}
\text{Z}=\Phi^{-1}(1-p),
\ee
where $\Phi^{-1}$ is the quantile function of a standard Gaussian distribution. Different tests are compared by evaluating their \emph{power}, namely their rate of true positives at the critical value $t_{\alpha}$
\begin{align}\label{eq:power}
&\alpha = P(t>t_{\alpha}|H_0),\\
&\textrm{power} = P(t\geq t_\alpha|H_1).
\end{align}
Given $\alpha$, the best test is the one maximising the power with a false positive rate at most equal to $\alpha$.

\subsection{The NPLM methodology}
NPLM is an approach to signal-agnostic hypothesis testing based on machine learning that aims at approximating the maximum-likelihood-ratio test as defined by Neyman and Pearson~\cite{Neyman:1933wgr}. It is based on the idea of introducing a local deformation of the reference distribution (as defined in eq.\eqref{n_density})
\be
n_w(x)=e^{f_w(x)}n(x|R),
\ee
with $\mathcal{F}=\{f_w\}$ a rich family of functions parametrised by $w$. In \cite{Letizia:2022xbe} and in this work we consider kernel methods, for which the function $f_w$ is expresses as the following weighted sum
\be\label{kernel_methods}
f_w(x)=\sum_{i=1}^{N} w_i k_\sigma (x,x_i),
\ee
with the parameters $w$ to be selected from data and $N=N_\scriptR+N_\scriptD$ the total number of data points. Specifically, we use a Gaussian kernel 
\be\label{rbf}
k_\sigma (x,x')=\exp\left(-\frac{||x-x'||^2}{2\sigma^2}\right),
\ee
where $\sigma$ is the kernel width, a hyperparameter. The resulting space of functions allows to approximate any continuous function given enough data. This approach is powerful but limited by large computational requirements. To solve this problem we use Falkon \cite{meanti2020kernel}, a modern solver for large-scale kernel methods which replaces eq.~\eqref{kernel_methods} with
\be\label{nystrom_km}
f_w(x)=\sum_{i=1}^{M} w_i k_\sigma (x,x_i),
\ee
where $\{\tilde{x}_1,...,\tilde{x}_M\}$ are called Nyström centres and are sampled uniformly at random from the input data, with $M$ a hyperparameter. The corresponding solution can be shown to be with high probability as accurate as the exact one (see \cite{rudi2021generalization} and references therein). In practice, the optimal parameters $\hat{w}$ are learned from data with a supervised classifier trained to separate $\scriptR$ from $\scriptD$ by minimising the following empirical risk
\be\label{reg_ERM}
\frac{1}{N}\sum_i\ell (y_i,f_w(x_i))+\lambda R(f_w),
\ee
based on a weighted logistic loss
\be
\ell(y,f_w(x))=(1-y)\frac{N(R)}{N_\scriptR} \log\left(1+e^{f_w (x)}\right)+y\log\left(1+e^{-f_w (x)}\right),
\ee
with $y=0$ if $x\in\scriptR$ and $y=1$ if $x\in\scriptD$. This loss can be shown (see \cite{Letizia:2022xbe}) to have the correct target function
\be
f_{\hat{w}}(x)\approx f^*(x)=\argmin_{f}\mathbb{E}\left[\ell(y,f(x))\right] = \log \frac{n_\true(x)}{n(x|R)}.
\ee
The second term in eq.~\eqref{reg_ERM} is a regularisation term
\be
R(f_w)=\sum_{ij} w_i w_j k_{\sigma}(x_i,x_j).
\ee
constraining the complexity of the model. The problem defined in eq.~\eqref{reg_ERM} is then solved  by an approximate Newton method, as discussed in detail in \cite{meanti2020kernel}. 

At the end of training, the model is evaluated in-sample on the whole dataset with the following metric
\be\label{nplm_statistic}
t_{\rm obs}(\scriptD)=-2\left[\frac{N(R)}{N_\scriptR}\sum_{x\in\scriptR}\left(e^{f_{\hat{w}}(x)}-1\right)-\sum_{x\in\scriptD} f_{\hat{w}}(x)\right],
\ee
which is derived from the extended likelihood-ratio (see \cite{barlow1990extended,DAgnolo:2018cun,Letizia:2022xbe}). To simplify the notation, we omit the dependence of eq.~\eqref{nplm_statistic} on the reference sample $\scriptR$ and on the learned parameters $\hat{w}$. This method allows to leverage the Neyman--Pearson approach to hypothesis testing with a data driven alternative hypothesis, without the need to specify it a priori. 
The connection between goodness-of-fit tests and the Neyman--Pearson construction at the core of NPLM was discussed earlier in \cite{Baker:1983tu} and more recently in \cite{Grosso:2023scl}.

\subsubsection{Model selection}\label{subsec:model_selection}
Falkon possesses three main hyperparameters: the number of centres $M$, the kernel width $\sigma$ and the regularisation parameter $\lambda$. These are tuned only on reference data to avoid biases toward specific anomalous features that might be present in the measurements $\scriptD$. Following \cite{Letizia:2022xbe,Grosso:2023ltd}, they are selected as follows:
\begin{itemize}
\item The Gaussian width $\sigma$ is selected as the 90th percentile of the pairwise distance among reference-distributed data points. Heuristics of this type are common for kernel methods, see for instance \cite{gretton2012kernel}.
\item To achieve optimal statistical bounds and preserve performance, the number of centres $M$ must be at least be of order $\sqrt{N}$, as discussed in~\cite{rudi2016more}.
Studies presented in \cite{Letizia:2022xbe} suggest that values close to the number of data points  ${N}_{\cal{D}}$ in the measurements work well but can be reduced for a faster training.
\item The regularisation parameter $\lambda$ is kept as small as possible while maintaining a stable training \cite{rudi2016more}.
\end{itemize}
As a consequence of these criteria, we consider the kernel width as the main hyperparameter that regulates the complexity of the model and sets the typical scale of the problem. Indeed, it is easy to show that if $\sigma$ is small the model tends to overfit while, if large, it behaves as a linear model.\footnote{It is however worth highlighting that, in general, all three hyperparameters act as regularisers.} In the context of two-sample testing, the specific choice of $\sigma$ has a crucial impact on the families of alternative hypotheses that are effectively explored by the test, as we will discuss in Sections \ref{sec:3} and \ref{sec:4} (see also \cite{schrab2023mmd,biggs2023mmdfuse}). 

\subsubsection{Single test at fixed hyperparameters}\label{subsubsec:single_nplm}
Given a particular set of hyperparameters $\theta^*=(M^*,\sigma^*,\lambda^*)$, a single test proceeds as follows. As a first step, the model is trained on the reference sample $\scriptR$ and the measurements $\scriptD$, returning the value of the observed test statistic $t_{\rm obs}=t_{\rm obs}(\scriptD)$, as given by eq.~\eqref{nplm_statistic}. Next, the distribution of the test statistic under the null hypothesis $p(t|H_0)$ is estimated empirically. There are different ways to do it. We consider here the scenario in which the reference model can be sampled at will via simulations. Therefore, we re-train the NPLM model from scratch on the reference sample $\scriptR$ and multiple ($N_{\rm toys}^{(H_0)}$) reference-distributed samples $\scriptD_i^{(R)}$, mimicking measurements in the absence of new physics. Each test returns a value $t_i=t(\scriptD_i^{(R)})$. The collection of test statistics $\{t_i\}_{i=1}^{N_{\rm toys}^{(H_0)}}$ is used to empirically estimate the p-value as (see \cite{north2002note})
\be\label{eq:emp_pvalue}
\hat{p}_{\rm obs} = \frac{1}{N_{\rm toys}^{(H_0)}+1} \left[\sum_{i=1}^{N_{\rm toys}^{(H_0)}} \mathbb{1}(t_i-t_{\rm obs})+1\right],
\ee
where $\mathbb{1}(x)$ is the Heaviside step function, which is zero when $x<0$ and one otherwise.
It is worth stressing that the result of the test is implicitly conditioned on the selected hyperparameters.

\section{Multiple tests for robust detection}\label{sec:3}

\subsection{The multiple testing problem}

In two-sample testing, one is interested in determining whether the null hypothesis that two samples are drawn from the same probability distribution can be rejected. The alternative hypothesis is the negation of the null and no assumption is made about how the data-generating distributions might differ. In practice, a specific test statistic has to be chosen to formulate a concrete procedure and this will in general bias the test towards specific hypotheses. For example, both the Kolmogorov-Smirnov and the Andreson-Darling tests \cite{darling1957kolmogorov} are viable options for a non-parametric test. However, the latter is more sensitive to discrepancies in the tails of the distributions. It is therefore logical to explore the possibility of conducting multiple tests to enhance the likelihood of detection.

The problem of multiple testing, also known as the \emph{look-elsewhere effect} in the HEP literature (see \cite{Gross:2010qma,Lyons_2008,Demortier:2007zz}), arises in this type of scenarios. Each individual test outputs a p-value. Naively, it would be ideal to simply retain the test returning the smallest p-value, associated with the highest detected degree of discrepancy. However it is not correct to simply compare the smallest p-value to the desired false positive rate $\alpha$. Indeed, it is crucial to take into account the fact that we are (at least implicitly) testing different hypotheses simultaneously, resulting in an increased possibility of having at least one false detection among the collection of considered tests (see~\cite{kuehl2000design}).
To address this problem, several methods to combine tests into a single meta-test have been explored in the literature. 
In common settings, all tests are designed to be sensitive to a specific signal of interest, i.e. they all share the same alternative hypothesis, and are applied to sets of independent measurements.
The reader can find in~\cite{heard2018choosing} an overview of the most common approaches based on combining p-values, and theoretical arguments on their optimality given a specific class of alternative hypothesis. 

In this work, we are interested in the case in which the tests are performed on the same set of measurements, hence with a potentially high degree of correlation. An example of this scenario can be found in~\cite{Rolke:2020stf}, in the context of common goodness-of-fit tests in one dimension.
Here, we employ multiple testing strategies to reduce the bias in the anomaly detection task caused by specific hyperparameter choices in the machine learning model powering the NPLM test.

\subsection{Designing multiple tests for NPLM}\label{subsec:design}
As discussed at the end of \ref{subsec:model_selection}, our multiple testing strategy focuses on the kernel width $\sigma$ in eq.~\eqref{rbf}, while the other hyperparameters of the kernel methods, $M$ and $\lambda$, are kept fixed. We proceed by choosing a set of unique values $\Sigma=\{\sigma_i\}_{i=1}^n$, defining the following set of tests
\be\label{eq:T}
T = \{t^{(\sigma_i)} | \sigma_i \in \Sigma \}_{i=1}^n .
\ee
In this regard, it is important to realise that values of $\sigma$ that are close will give rise to highly correlated tests, while far apart values will result in less correlated tests. 

Additionally, since the test considered here is based on a learning model, it will be highly adaptive to the data, potentially increasing correlation among tests. To ensure robust performances across different anomalous scenarios and decrease correlation, few well-separated values of $\sigma$ are preferable. Following a standard practice in kernel methods (see for instance \cite{gretton2012kernel}) and similarly to the original proposal presented in \cite{Letizia:2022xbe}, we select them as percentiles of the distribution of the pairwise distance in a set of reference-distributed data points, after proper feature rescaling. This provides an estimate of the relevant scales in the problem. However, it might be beneficial to also include larger values to consider possible long-range effects. The number of tests $n=|\Sigma|$ is a free parameter of the algorithm. In choosing it, one should keep in mind that performing an extensive number of tests is computationally more demanding, although this cost can in principle be amortised with adequate distributed computing strategies. On the other hand, this could ultimately have a negative impact on the sensitivity if, as $n$ grows, the rate of false positives increases faster than the rate of true positives. 

\subsection{Aggregation methods}\label{subsec:aggreg}
We explore various options to combine tests based on the existing literature, and discuss the benefits and disadvantages of each of them given the design choices outlined in Section~\ref{subsec:design}. Specifically, we consider the following meta-test statistics:
\begin{description}
\item[min-$p$]
Introduced in \cite{tippett1931methods}, the meta-test statistic is defined as the smallest individual p-value
\be
p_{\rm min} = -\log \min_{\sigma\in\Sigma} p^{(\sigma)}.
\ee
\item[prod-$p$] Following \cite{fisher1970statistical}, the meta-test statistic is defined as the log-scaled product of the individual p-values
\be
p_{\rm prod} = -\sum_{\sigma\in\Sigma} \log p^{(\sigma)}.
\ee
\item[avg-$p$] Similarly to \cite{edgington1972additive}, the individual p-values are averaged as
\be
p_{\rm avg} = -\frac{1}{n}\sum_{\sigma\in\Sigma} p^{(\sigma)}.
\ee
\item[smax-$t$]
Inspired by \cite{biggs2023mmdfuse}, the test statistics are combined directly via the following smooth maximum function
\be
    t_{\rm smax} = T \log \frac{1}{n}\sum_{\sigma\in\Sigma} e^{t^{(\sigma)}/T},
\ee
where $T\in\mathbb{R}_{>0}$ plays the role of a temperature. For $T\rightarrow 0$ it corresponds to the maximum, while for $T\rightarrow \infty$ it reduces to the arithmetic mean (see \cite{asadi2017alternative}). We fix $T=1$, unless specified otherwise. 
\end{description}

Choosing the optimal method without prior knowledge on the type of signal potentially in the measured data is generally not a solvable problem for composite hypotheses. However, 
the specificities of these tests can be used as a guide to isolate the most promising options.

The avg-$p$ method assigns uniform weights to all the tests. This is generally a good choice if the tests are expected to perform similarly.
This is not necessarily the case for the NPLM set of tests, since the choice of $\Sigma$ is made such that the overlap between families of alternatives is small.

The log-scaled product of p-values (prod-$p$) allows to direct the combination focus toward the smallest p-values.
This can be a good choice if a subgroup of the tests performs well relative to the others, as it allows to enhance their contributions to the sum. 

The minimum over p-values (min-$p$) is intuitively the best solution if a specific test is expected to perform significantly better than the others.

The typical values of the NPLM test statistic strongly depend on the complexity of the model. In particular, for any given set of data, $t^{(\sigma_1)}>t^{(\sigma_2)}$ if $\sigma_1<\sigma_2$, as also observed in previous studies (\cite{DAgnolo:2018cun,DAgnolo:2019vbw,Letizia:2022xbe}).
The result of combining tests via smax-$t$ is therefore equivalent to selecting the test statistic with smallest value of $\sigma$. Therefore, we do not expect this strongly biased strategy to work well in our study.

Estimating the level of correlation among NPLM tests is thus crucial to identify the best aggregation method.
A signal-agnostic strategy to address this task is to inspect the pairwise correlation under the null hypothesis, i.e. when detecting statistical fluctuations in background-only samples $\scriptD_i^{(R)}$. We will give practical examples within the scope of our numerical experiments in Section~\ref{sec:4}.

\section{Numerical results}\label{sec:4}
\subsection{Methodology}
This section is dedicated to comparing the different approaches to multiple testing considered in this work and outlined in Section~\ref{subsec:aggreg}. We utilize three benchmarks from the high-energy physics literature on signal-agnostic searches and anomaly detection \cite{DAgnolo:2018cun,DAgnolo:2019vbw,Grosso:2023scl,Kasieczka:2021xcg} with minor modifications. Each benchmark is defined by a reference distribution, characterising the null hypothesis, and different types of new physic signals, characterising different alternative hypotheses. When possible, we vary certain parameters to alter the alternative, such as the width of a resonance or the number of signal events, to explore more diverse scenarios and increase the validity of our study. 

For each benchmark and each value of the kernel width $\sigma$ in the predefined set $\Sigma$, we estimate the distribution of the NPLM test under the null hypothesis $p(t^{(\sigma)}|H_0)$ as outlined in Section \ref{subsubsec:single_nplm}. We also characterise the distribution under each of the alternative hypotheses $p(t^{(\sigma)}|H_1)$ by computing the test independently on a set of 
$N_{\rm toys}^{(H_1)}$ samples for each new physics signal. The $N_{\rm toys}^{(H_{0/1})}$ samples are drawn from the true data-generating distributions whenever possible. Alternatively, following a bootstrap-based approach, they are sampled with replacement from a larger set of events.

To calculate the meta-tests min-$p$, prod-$p$ and avg-$p$, for each $\sigma\in\Sigma$ we compute the $p$-value $p^{(\sigma)}$ associated with each toy, either drawn from the reference or from the alternative hypotheses, according to eq.~\eqref{eq:emp_pvalue}. 
For each reference-distributed sample, we pay attention not to include the test of interest in the null distribution, and thus we compute its $p$-value with respect to the remaining $N_{\rm toys}^{(H_0)}-1$ tests.

As outlined in Section~\ref{sec:2}, we evaluate the meta-tests by computing an their powers on each benchmark and alternative hypothesis. To do so, we select two values for the desired rate of false positives $\alpha$, given as Z-scores according to eq.~\eqref{eq:zscore}. We take $Z_\alpha=3$, namely the threshold used in HEP to report evidence of a signal, and $Z_\alpha=2$. To convert each of these numbers to critical values of the test statistic $t_\alpha$, we consider the $N_{toys}^{(H_0)}$ values of the test statistic computed under the null hypothesis and we take the highest value corresponding to an empirical quantile not larger that $1-\alpha$. We adopt this empirical procedure because we do not know the distribution of the test statistic under the null hypothesis analytically. It is a conservative approach that results in a true rate of false positives $\alpha'\leq\alpha$. Similarly to eq.~\eqref{eq:emp_pvalue}, we then estimate the power, given by eq.~\eqref{eq:power}, empirically using the following sum
\be\label{emp_power}
\widehat{\textrm{power}}_\alpha=\frac{1}{N_{\rm toys}^{(H_1)}}\sum_{i=1}^{N_{\rm toys}^{(H_1)}} \mathbb{1}(t_i-t_\alpha).
\ee

\subsection{Datasets and hyperparameters}
\subsubsection{EXPO-1D}\label{subsubsec:expo1d}
In this univariate benchmark (see also \cite{DAgnolo:2018cun,Letizia:2022xbe,Grosso:2023scl}), we consider a reference model given by an energy spectrum that decays exponentially, described by the following density
\be
n(x|{\rm R})=N(R) e^{-x} \,,
\ee
where the expected number events in the reference hypothesis is $N(R)=2000$. The reference sample is composed of $N_\scriptR=100\,N(R)$ events.
We consider the following parametrised alternative hypothesis
\be
n_{\rm t
rue}(x)=n(x|{\rm R})+N(S) \frac{1}{\sqrt{2\pi}\sigma_{\rm NP}} \exp\left[-\frac{(x-\bar{x}_{\rm NP}^2)}{2\sigma_{\rm NP}^2}\right],
\ee
representing a Gaussian peak with mean $\bar{x}_{\rm NP}$ and standard deviation $\sigma_{\rm NP}$, on top of the reference background. In our tests, we vary both parameters and the average number of injected new physics events $N(S)$ to establish the performance of the method.

\begin{figure}[H]
    \centering
    \includegraphics[width=1\linewidth]{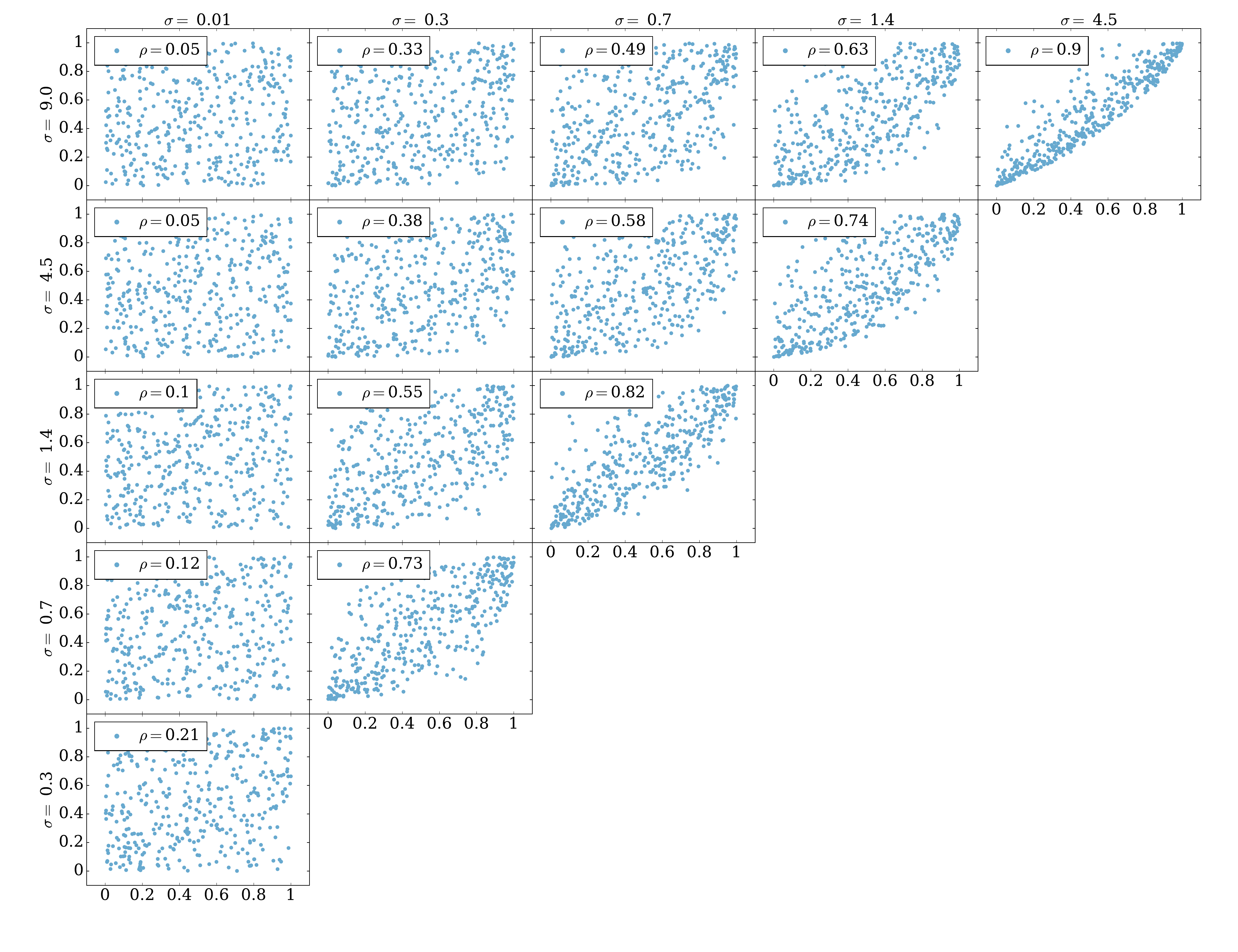}
    \caption{\textbf{EXPO-1D --} Corner plots showing correlations between the p-values obtained from different tests in the background-only hypothesis. The Pearson's correlation ($\rho$) is reported in the legend.}
    \label{fig:1d-correlations}
\end{figure}
\paragraph{Hyperparameters} 
Being an illustrative benchmark, we select values for $M$ and $\lambda$ that result in faster training times than those reported in \cite{Grosso:2023scl}. We use $M=1000$, $\lambda=10^{-6}$ and $\Sigma=\{0.01,0.3,0.7,1.4,4.5, 9\}$. The first five values correspond to the 0.01, 0.25, 0.50, 0.75 and 0.99 quantiles; the last value is chosen as twice the value of the 0.99 quantile.
Correlation among tests can be studied and estimated in a signal-agnostic way by inspecting the pairwise correlations under the null hypothesis, as depicted in Figure~\ref{fig:1d-correlations}. Here, the panels closer to the diagonal show the correlation between tests that are closer in the $\sigma$ space. As anticipated in Section~\ref{sec:3}, the correlation is higher for nearby tests.

\subsubsection{MUMU-5D}
This five dimensional dataset (introduced in \cite{DAgnolo:2019vbw}) is composed of simulated LHC collision events producing two opposite charged muons in the final state ($pp\rightarrow\mu^+\mu^-$)  at a center-of-mass energy of 13 TeV. \footnote{Data available at \url{https://zenodo.org/record/4442665}} The features are the transverse momenta and pseudorapidities of the two muons, and their relative azimuthal angle, i.e., $x=[p_{T1},p_{T2},\eta_1,\eta_2,\Delta\phi]$.   We consider two types of new physics contributions: the first one is a new vector boson ($Z'$) for which we study different mass values ($m_{Z'}=180,200$ and 600 GeV); the second one is a non-resonant signal obtained by adding a four-fermion contact interaction to the Standard Model Lagrangian for which the Wilson coefficient $c_W$ determines the coupling strength. 
We fix $N(R)=2\times 10^4$ expected events in the reference hypothesis and the size of the reference sample is $N_\scriptR=10^5$. Also in this case, we vary the number of expected signal events $N(S)$.
\paragraph{Hyperparameters} 
We selected $M=10^4$, $\lambda=10^{-6}$ and $\Sigma=\{0.31,1.19,1.79,2.49,4.23, 8.0\}$. The first five values correspond to the 0.01, 0.25, 0.50, 0.75 and 0.99 quantiles; the last value is chosen as approximately twice the 0.99 quantile.

\subsubsection{LHCO-6D}
The LHC Olympic dataset is a widely utilised benchmark for resonant anomaly detection proposed as a challenge by~\cite{Kasieczka:2021xcg}. The dataset, available on Zenodo (~\cite{lhco4536377}), consists of LHC collision events with two jets in the final state. 
The Standard Model background consists of QCD events while the signal is modelled as a resonant $W'$ decaying into two massive particles $X$ and $Y$, with $X\rightarrow qq$ and $Y\rightarrow qq$. The $W'$, $X$, and $Y$ masses are 3.5 TeV, 500 GeV and 100 GeV respectively.
Both constituents level and jet level information is provided for each event. In this application we focus on six high level observables describing the dijet system: the dijet invariant mass, the mass of the leading jet, the difference between the two jets masses, the angular separation between the two jets, and the 2-subjettiness ratios for both jets ($\tau_{2,1}^{J1}$ and $\tau_{2,1}^{J2}$).
Events are required to have at least one $\mathrm{R} = 1.0$, pseudrapidity $\left| \eta \right| < 2.5 $, and transverse momentum $p_\mathrm{T}^{J} > 1.2$~TeV.
Since most of the applications concerning this dataset rely on a bump-hunt approach with sliding window on the dijet invariant mass, we focus our test on one single mass window, corresponding to the signal region ($3.1\le m_{J1,J2}\le 3.7$ TeV).
In this selection, the expected number of background events is approximately 121000, on top of which we inject 333 signal events.

\paragraph{Hyperparameters} 
We select $M=853$, $\lambda=10^{-6}$ and $\Sigma=\{1.2,2.5,3.2,3.9,5.1, 6.1, 12.2\}$. The first five values correspond to the 0.01, 0.25, 0.50, 0.75 and 0.99 quantiles; the last value is chosen as twice the 0.99 quantile.

\subsection{Results}

Tables~\ref{tab:1d-z3}--\ref{tab:lhco} summarise the power of the various meta-tests described in Section~\ref{sec:3} for all the benchmarks in this study. They report the probability of observing a Z-score greater or equal to $3$, in other words the chances of finding evidence for the signal, and the probability of observing a Z-score greater or equal to $2$. In the upper part of the tables, we show the sensitivity of the NPLM test for each individual value in $\Sigma$. In the middle part we show the performance of the standard NPLM approach presented in \cite{Letizia:2022xbe}. In the bottom part we report the performance of the various aggregation strategies introduced in Section~\ref{subsec:aggreg}.
All entries in the table are endowed with uncertainties computed as the 68\% Clopper-Pearson~\cite{10.1093/biomet/26.4.404} confidence interval.

\begin{figure}[h]
     \includegraphics[width=0.4\linewidth]{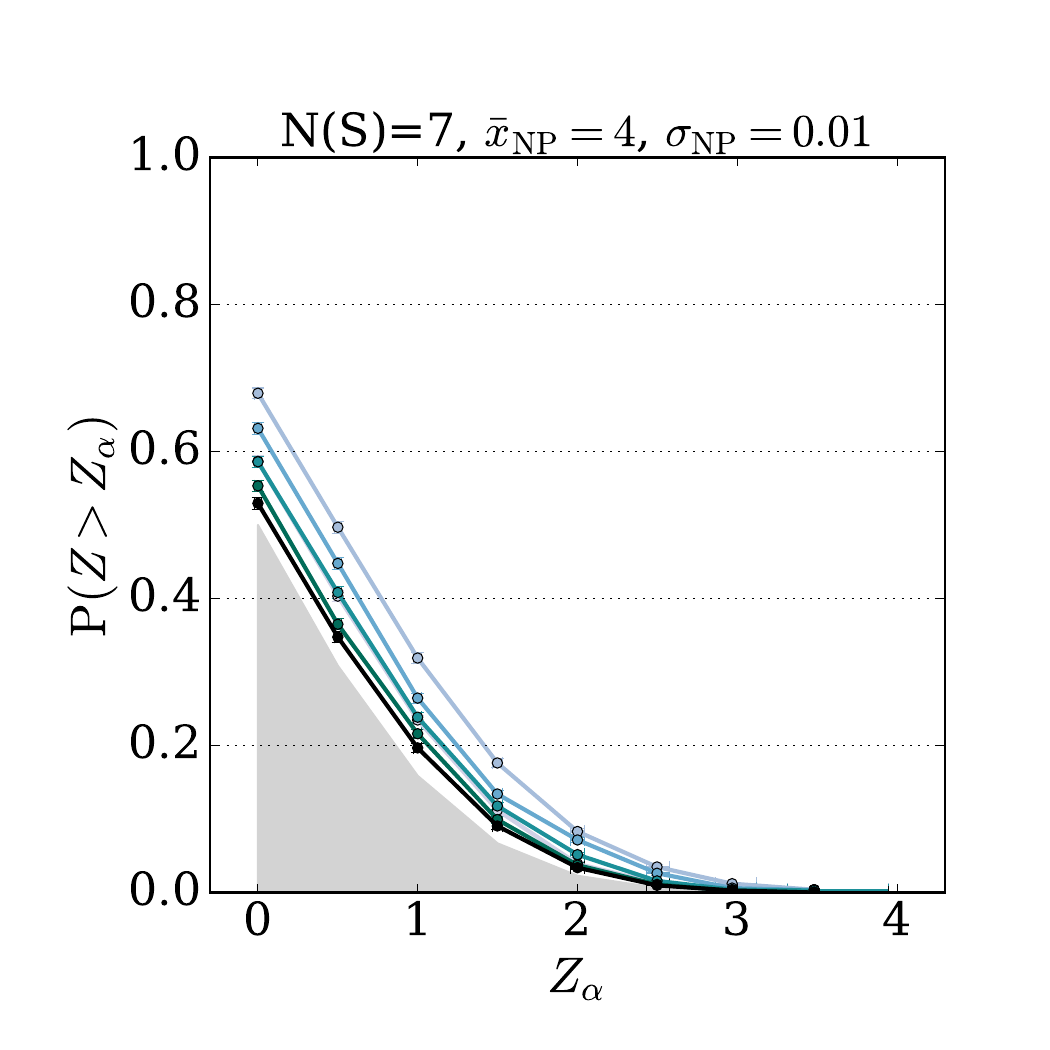}
     \includegraphics[width=0.56\linewidth]{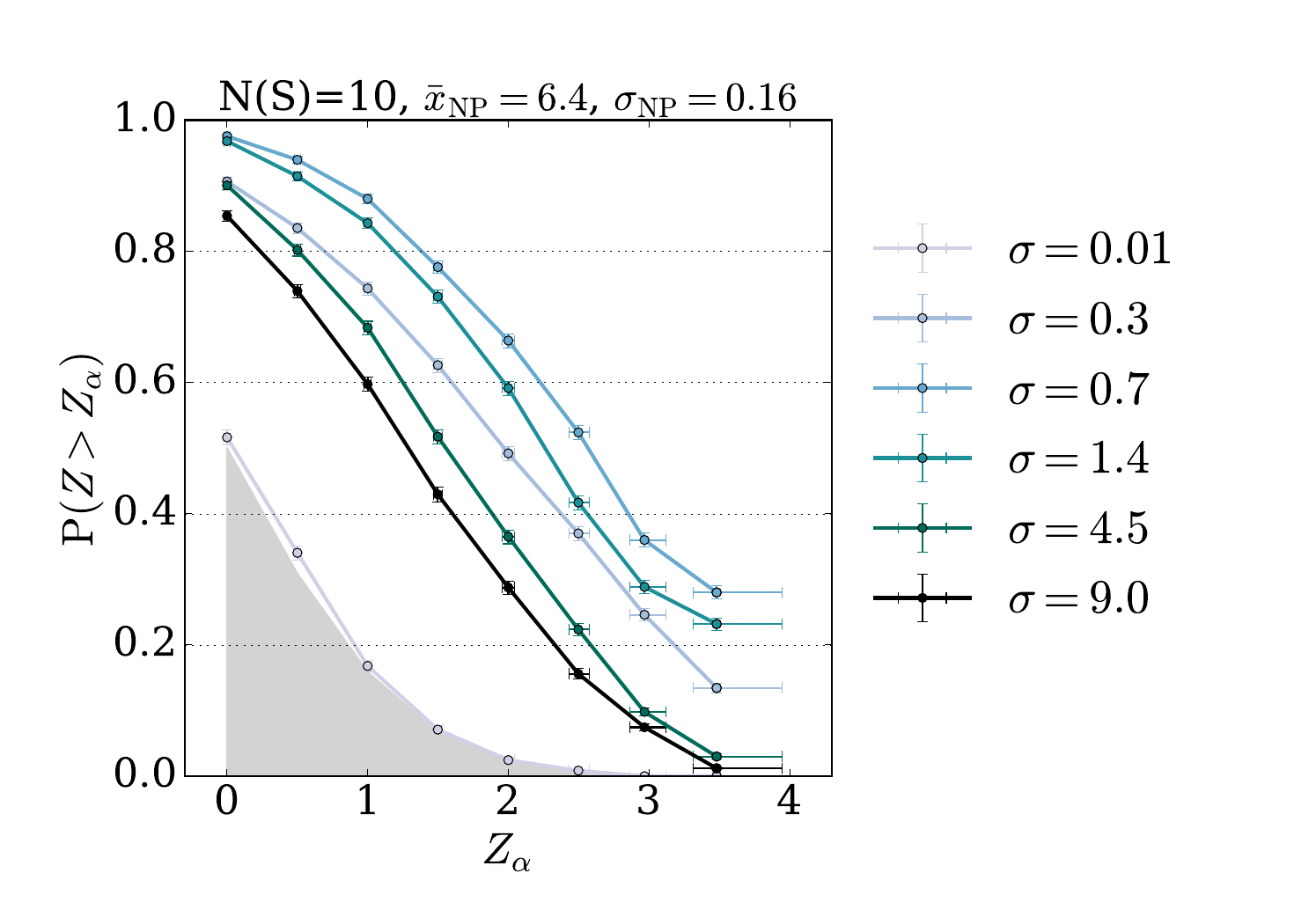}
     \caption{\textbf{EXPO 1D --} Illustrative examples of power curves for NPLM tests performed with different choices of $\sigma$. The left hand panel shows the power curve for a narrow signal, corresponding to the first columns in Table~\ref{tab:1d-z3}; the right hand panel shows the power curves for a signal in the tail (forth column in Table~\ref{tab:1d-z3}).}
     \label{fig:1d-expo-plots}
 \end{figure}

\begin{figure}[H]
    \centering
    \includegraphics[width=0.4\linewidth]{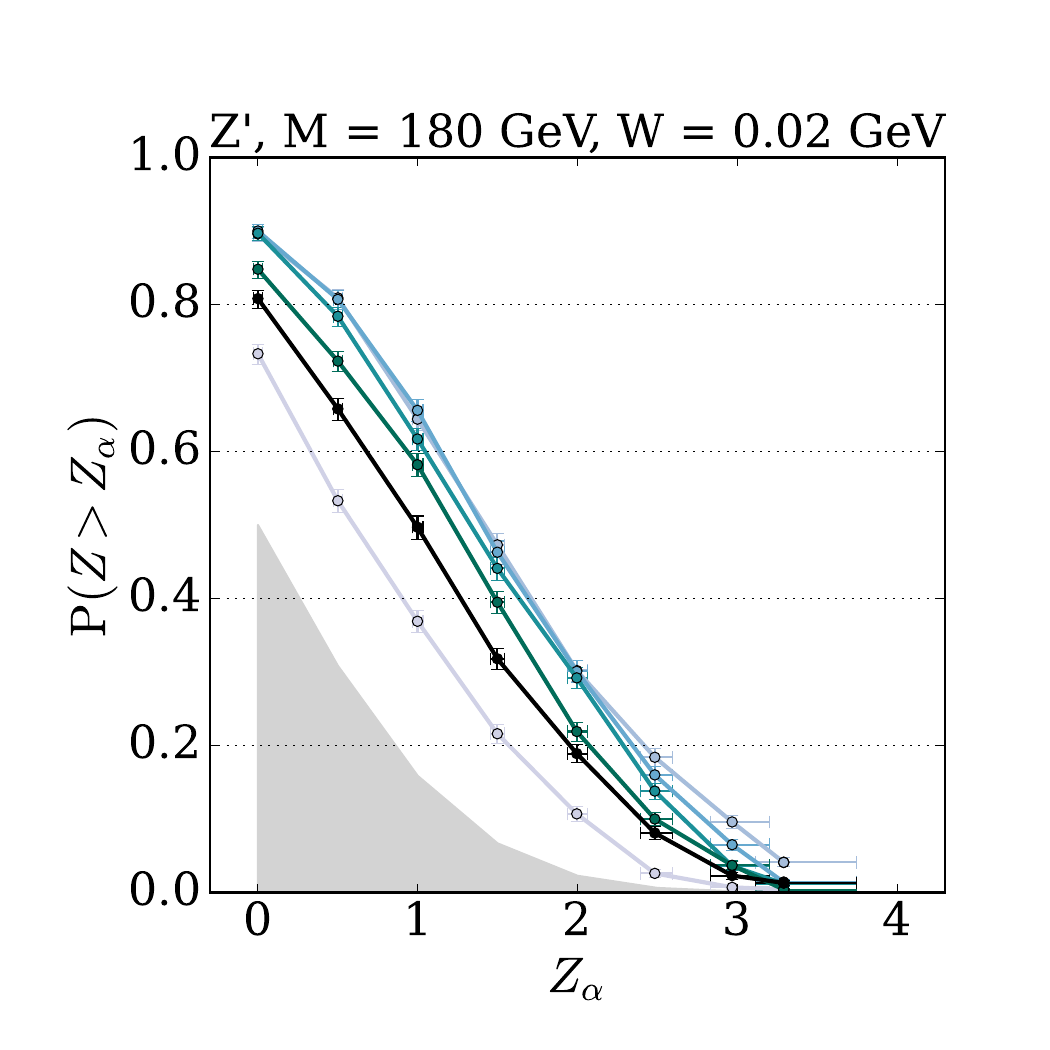}
    \includegraphics[width=0.56\linewidth]{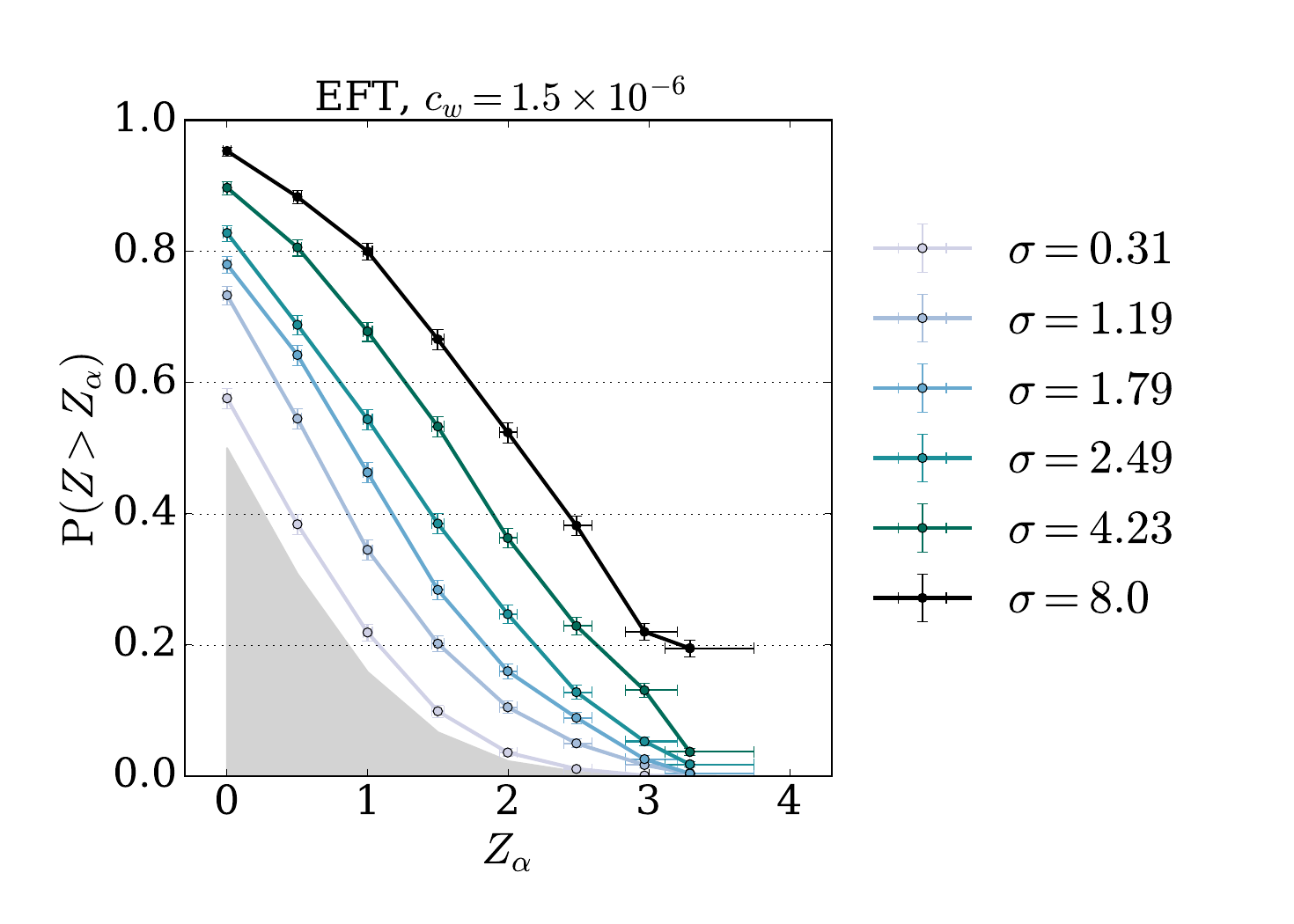}
    \caption{\textbf{MUMU-5D.} Power curves for different choices of $\sigma$.}
    \label{fig:mumu-p-comparison}
\end{figure}

The results obtained with single values of $\sigma$ highlight the dependency of the NPLM test outcome on the specific choices of kernel width and signal benchmark. Narrow peaks, like the one reported in the first columns of Tables~\ref{tab:1d-z3} and \ref{tab:mumu-z3}, are better detected by small values of $\sigma$, whereas large values are preferable to detect broad peaks, like the one reported in the third columns.
We report in Figures~\ref{fig:1d-expo-plots},\ref{fig:mumu-p-comparison} and \ref{fig:lhco} examples of the full power curve of each individual test, showing how the sensitivity changes according to $\sigma$ and follows different trends depending on the signal.

Our studies show that it is beneficial to combine multiple tests. With the exception of smax-$t$, that corresponds to systematically selecting the test with the smallest width, the other methods return powers that are comparable with or larger than the original kernel-based NPLM proposal in \cite{Letizia:2022xbe} and often competitive with the best overall test, which would be hard to identify a priori in real analyses. We observe that min-$p$ is the most robust choice across multiple signal scenarios. Indeed, it is the one that gives the best results in most cases and when it does not, its failure is not catastrophic. The advantage of using min-$p$ becomes particularly evident for instances of the test that are harder to occur. In this cases, and specifically if there is one test in the set $\Sigma$ with a significantly better induced bias
, min-$p$ is a good method. We also observe that prod-$p$ gives good results if a subset of tests performs similarly well as in the third column of Table~\ref{tab:1d-z3}. The method avg-$p$ is instead performing well when the there is not a strong separation between the performance of the individual tests, as shown in Table~\ref{tab:lhco}. These results confirm the intuitions discussed in Section~\ref{subsec:aggreg}.

\begin{table}[H]
    \centering
    \scalebox{0.99}{
    \begin{tabular}{c|ccccc}
    N(S)                & 7 & 18 & 13 & 10 & 90 \\
    $\bar x_{\rm NP}$            & 4 & 4 & 4 &6.4 &1.6\\
    $\sigma_{\rm NP}$   & 0.01 & 0.16 & 0.64 &0.16 &0.16\\
    \midrule
    \midrule
    $\sigma=0.01$
        & $0.0028\pm0.0008$
        & $0.0010\pm0.0006$
        & $0.0005\pm0.0004$
        & $0.0001\pm0.0001$
        & $0.029\pm0.004$\\
    $\sigma=0.3$
        & $\boldsymbol{0.012 \pm 0.002}$
        & $0.107\pm0.007$
        & $0.008\pm0.002$
        & $0.246\pm0.009$
        & $0.65\pm0.01$\\
    $\sigma=0.7$
        & $0.006\pm0.001$                 
        & $\boldsymbol{0.123\pm0.007}$
        & $\boldsymbol{0.011\pm0.002}$
        & $\boldsymbol{0.36\pm0.01}$
        & $\boldsymbol{0.70\pm0.01}$\\
    $\sigma=1.4$
        & $0.004\pm0.001$                
        & $0.078\pm0.006$
        & $\boldsymbol{0.012\pm0.002}$
        & $0.29\pm0.01$
        & $0.54\pm0.01$\\
    $\sigma=4.5$
        & $0.0023\pm0.0007$                
        & $0.020\pm0.003$
        & $\boldsymbol{0.011\pm0.002}$
        & $0.098\pm0.007$
        & $0.28\pm0.01$\\
    $\sigma=9.0$
        & $0.0028\pm0.0008$                
        & $0.018\pm0.003$
        & $\boldsymbol{0.012\pm0.002}$
        & $0.075\pm0.006$
        & $0.24\pm0.01$\\
    \midrule
    $\sigma=2.3$
        & $0.0023\pm 0.0007$              
        & $0.044\pm0.005$
        & $0.013\pm0.002$
        & $0.028\pm0.004$
        & $0.36\pm0.01$\\
    \midrule
    min-$p$ 
        & $\boldsymbol{0.008\pm0.001}$
        & $\boldsymbol{0.103\pm0.007}$
        & $0.007\pm0.002$
        & $\boldsymbol{0.32\pm0.01}$ 
        & $\boldsymbol{0.66\pm0.01}$\\
    prod-$p$ 
        & $0.005\pm0.001$
        & $0.083\pm0.006$
        & $\boldsymbol{0.012\pm0.002}$ 
        & $0.26\pm0.01$
        & $0.65\pm0.01$ \\
     
    avg-$p$ 
        & $0.006\pm0.001$
        & $0.049\pm0.005$
        & $0.011\pm0.002$
        & $0.068\pm0.006$ 
        & $0.50\pm0.01$\\
       
    smax-$t$ 
        & $0.0028\pm0.0008$
        & $0.0010\pm0.0006$
        & $0.0005\pm0.0004$
        & $0.0001\pm0.0001$
        & $0.029\pm0.004$\\ 
    \end{tabular}}
    \caption{\textbf{EXPO 1D --} probability of observing $Z\geq3$.}
    \label{tab:1d-z3}
\end{table}
\begin{table}[H]
    \centering
    \scalebox{0.99}{
    \begin{tabular}{c|ccccc}
    N(S)                & 7 & 18 & 13 & 10  & 90 \\
    $\bar x_{\rm NP}$   & 4 & 4  & 4  & 6.4 & 1.6\\
    $\sigma_{\rm NP}$   & 0.01 & 0.16 & 0.64 &0.16&0.16\\
    \midrule
    \midrule
    $\sigma=0.01$
        & $0.039\pm0.003$
        & $0.046\pm0.005$
        & $0.023\pm0.003$
        & $0.025\pm0.003$
        & $0.25\pm0.01$\\
    $\sigma=0.3$
        & $\boldsymbol{0.083 \pm 0.004}$
        & $0.35\pm0.01$
        & $0.053\pm0.005$
        & $0.49\pm0.01$
        & $0.881\pm0.008$\\
    $\sigma=0.7$
        & $0.072\pm0.004$                 
        & $\boldsymbol{0.37\pm0.01}$
        & $0.076\pm0.006$
        & $\boldsymbol{0.66\pm0.01}$
        & $\boldsymbol{0.913\pm0.007}$\\
    $\sigma=1.4$
        & $0.052\pm0.003$                
        & $0.28\pm0.01$
        & $\boldsymbol{0.082\pm0.006}$
        & $0.59\pm0.01$
        & $0.820\pm0.009$\\
    $\sigma=4.5$
        & $0.037\pm0.003$                
        & $0.166\pm0.008$
        & $0.080\pm0.006$
        & $0.37\pm0.01$
        & $0.63\pm0.01$\\
    $\sigma=9.0$
        & $0.0304\pm0.003$                
        & $0.121\pm0.007$
        & $0.074\pm0.006$
        & $0.29\pm0.01$
        & $0.58\pm0.01$\\
    \midrule
    $\sigma=2.3$
        & $0.039\pm 0.003$              
        & $0.207\pm0.009$
        & $0.079\pm0.006$
        & $0.48\pm0.01$
        & $0.69\pm0.01$\\
    \midrule
    min-$p$ 
        & $0.063\pm0.004$
        & $0.31\pm0.01$
        & $0.066\pm0.006$
        & $\boldsymbol{0.58\pm0.01}$
        & $0.877\pm0.007$\\
    prod-$p$ 
        & $\boldsymbol{0.065\pm0.004}$
        & $\boldsymbol{0.32\pm0.01}$
        & $\boldsymbol{0.085\pm0.006}$
        & $\boldsymbol{0.58\pm0.01}$ 
        & $\boldsymbol{0.897\pm0.007}$\\
    avg-$p$ 
        & $0.063\pm0.004$
        & $0.230\pm0.009$
        & $0.082\pm0.006$
        & $0.34\pm0.01$ 
        & $0.835\pm0.009$\\
    smax-$t$ 
        & $0.039\pm0.003$
        & $0.046\pm0.005$
        & $0.023\pm0.003$
        & $0.025\pm0.003$ 
        & $0.25\pm0.01$\\
    \end{tabular}}
    \caption{\textbf{EXPO 1D --} Probability of observing $Z\geq2$.}
    \label{tab:1d-z2}
\end{table}	 

\begin{table}[H]
    \centering
    \scalebox{0.99}{
    \begin{tabular}{l|cccccc}
    test
            & Z' M = 180 GeV 
            & Z' M = 300 GeV
            & Z' M = 600 GeV
            & EFT
            \\
            & W = 0.02 GeV
            & W = 15 GeV
            & W = 30 GeV
            & $c_w=1.5\times 10^{-6}$
            \\
    \midrule
    \midrule
    $\sigma=0.31$
    & $0.007\pm0.003$
    & $0.004\pm0.002$ 
    & $0.0010\pm0.0008$ 
    & $0.0010\pm0.0008$\\
    
    $\sigma=1.19$
    & $\boldsymbol{0.096\pm0.009}$
    & $0.10\pm0.01$ 
    & $0.006\pm0.002$ 
    & $0.017\pm0.004$ \\
    
    $\sigma=1.79$
    & $0.065\pm0.008$
    & $0.11\pm0.01$ 
    & $0.012\pm0.003$ 
    & $0.026\pm0.005$\\
    
    $\sigma=2.49$
    & $0.036\pm0.006$          
    & $0.11\pm0.01$ 
    & $0.027\pm0.005$ 
    & $0.053\pm0.007$\\
    
    $\sigma=4.23$
    & $0.037\pm0.006$
    & $\boldsymbol{0.13\pm0.01}$ 
    & $\boldsymbol{0.066\pm0.008}$ 
    & $0.13\pm0.01$\\

    $\sigma=8.0$
    & $0.023\pm0.004$
    & $0.068\pm0.008$ 
    & $0.056\pm0.007$ 
    & $\boldsymbol{0.22\pm0.01}$\\
    \midrule
    $\sigma=3.0$
    & $0.031\pm0.005$
    & $0.13\pm0.01$ 
    & $0.044\pm0.006$ 
    & $0.092\pm0.009$\\
    \midrule
    min-$p$ 
        & $0.065\pm0.008$
        & $0.16\pm0.01$ 
        & $\boldsymbol{0.057\pm0.007}$
        & $\boldsymbol{0.23\pm0.01}$\\
    prod-$p$ 
        & $0.089\pm0.009$
        & $\boldsymbol{0.18\pm0.01}$ 
        & $0.028\pm0.005$ 
        & $0.083\pm0.009$\\
    avg-$p$ 
        & $\boldsymbol{0.14\pm0.01}$
        & $0.15\pm0.01$
        & $0.035\pm0.006$ 
        & $0.098\pm0.009$\\
    smax-$t$ 
        & $0.007\pm0.003$
        & $0.004\pm0.002$
        & $0.0010\pm0.0008$ 
        & $0.0010\pm0.0008$\\
    \end{tabular}}
    \caption{\textbf{MUMU 5D --} Probability of observing $Z\geq 3$.}
    \label{tab:mumu-z3}
\end{table}	

\begin{table}[H]
    \centering
    \scalebox{0.99}{
    \begin{tabular}{l|cccccc}
    test
            & Z' M = 180 GeV 
            & Z' M = 300 GeV
            & Z' M = 600 GeV
            & EFT
            \\
            & W = 0.02 GeV
            & W = 15 GeV
            & W = 30 GeV
            & $c_w=1.5\times 10^{-6}$
            \\
    \midrule
    \midrule
    $\sigma=0.31$
    & $0.11\pm0.01$
    & $0.042\pm0.006$ 
    & $0.023\pm0.005$ 
    & $0.036\pm0.006$\\
    
    $\sigma=1.19$
    & $\boldsymbol{0.30\pm0.01}$
    & $0.35\pm0.02$ 
    & $0.047\pm0.007$ 
    & $0.11\pm0.01$\\
    
    $\sigma=1.79$
    & $\boldsymbol{0.30\pm0.01}$
    & $0.41\pm0.02$ 
    & $0.11\pm0.01$ 
    & $0.16\pm0.01$\\
    
    $\sigma=2.49$
    & $0.25\pm0.01$          
    & $\boldsymbol{0.42\pm0.02}$ 
    & $0.19\pm0.01$ 
    & $0.25\pm0.01$\\
    
    $\sigma=4.23$
    & $0.23\pm0.01$
    & $0.41\pm0.02$ 
    & $0.25\pm0.01$ 
    & $0.32\pm0.01$\\
    $\sigma=8.0$
    & $0.19\pm0.01$
    & $0.31\pm0.01$ 
    & $\boldsymbol{0.29\pm0.01}$ 
    & $\boldsymbol{0.52\pm0.02}$\\
    \midrule
    $\sigma=3.0$
    & $0.25\pm0.02$
    & $\boldsymbol{0.42\pm0.02}$ 
    & $0.24\pm0.02$ 
    & $0.30\pm0.02$\\
    \midrule
    min-$p$ 
        & $0.32\pm0.01$
        & $0.47\pm0.02$
        & $\boldsymbol{0.28\pm0.01}$
        & $\boldsymbol{0.53\pm0.02}$\\
    prod-$p$ 
        & $\boldsymbol{0.38\pm0.02}$
        & $\boldsymbol{0.53\pm0.02}$ 
        & $0.23\pm0.01$ 
        & $0.38\pm0.02$\\
    avg-$p$ 
        & $0.37\pm0.02$
        & $0.46\pm0.02$
        & $0.18\pm0.01$ 
        & $0.31\pm0.01$\\
    
    smax-$t$ 
        & $0.11\pm0.01$
        & $0.042\pm0.006$
        & $0.023\pm0.005$ 
        & $0.036\pm0.006$\\
    \end{tabular}}
    \caption{\textbf{MUMU 5D --} Probability of observing $Z\geq 2$. }
    \label{tab:mumu-z2}
\end{table}	

\begin{minipage}{0.51\linewidth}
\begin{figure}[H]
    \centering
    \includegraphics[width=1.07\linewidth]{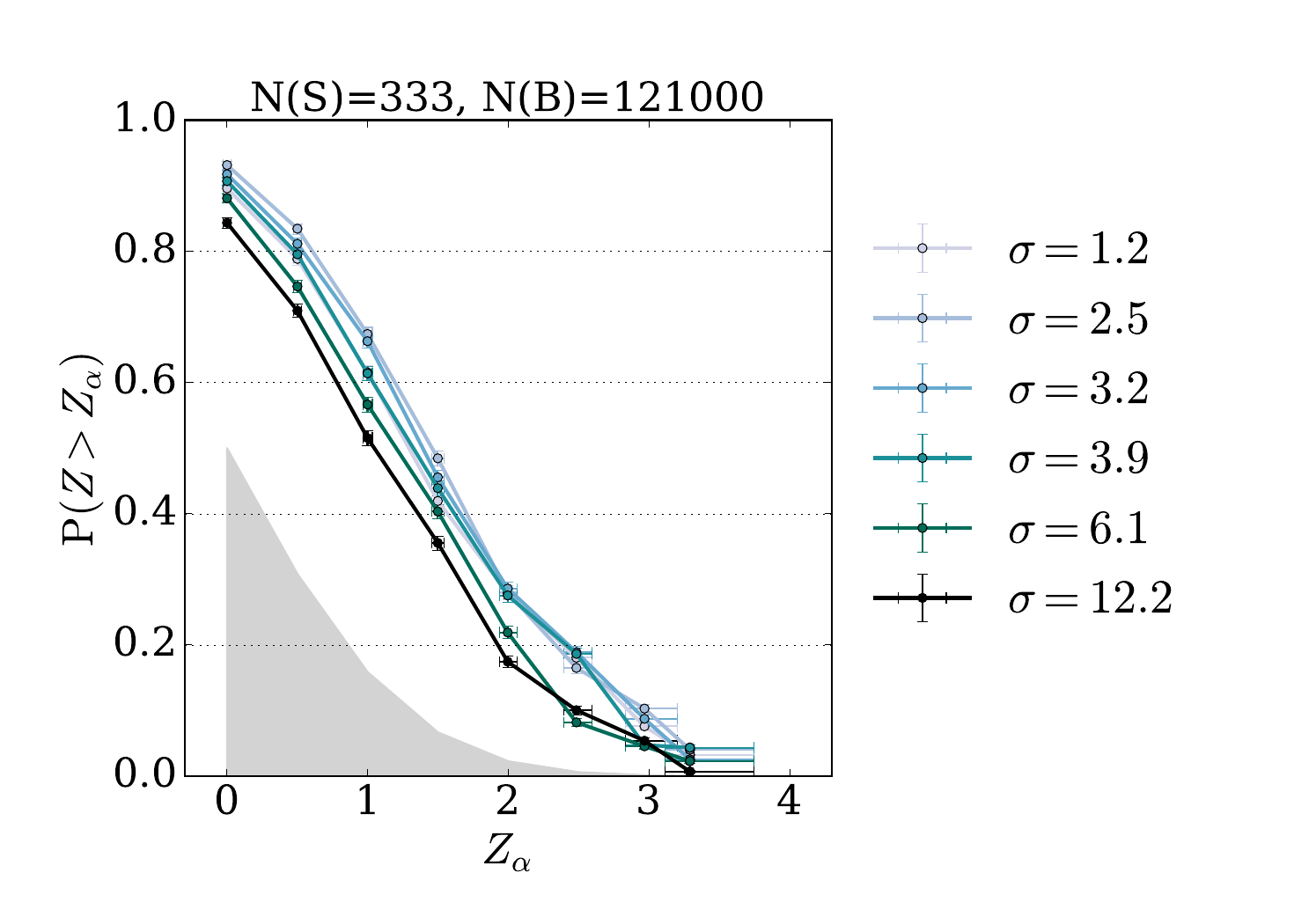}
    \caption{\textbf{LHCO-6D.} Power curves for different choices of $\sigma$ (shades of green lines), compared with the one for the $\min$-$p$ aggregation (black line).}
    \label{fig:lhco}
\end{figure}
\end{minipage}\hfill
\begin{minipage}{0.42\linewidth}
\begin{table}[H]
    \centering
    \scalebox{0.99}{
    \begin{tabular}{l|cc}
    test&$Z\ge2$&$Z\ge3$\\
    \midrule
    \midrule
    $\sigma=1.2$
        & $0.28\pm0.01$
        & $0.0763\pm0.006$\\
    $\sigma=2.5$
        & $0.28\pm0.01$
        & $\boldsymbol{0.103\pm0.007}$\\
    $\sigma=3.2$
        & $\boldsymbol{0.29\pm0.01}$
        & $0.088\pm0.006$\\
    $\sigma=3.9$              
        & $0.28\pm0.01$
        & $0.048\pm0.005$\\
    $\sigma=6.1$             
        & $0.219\pm0.009$
        & $0.046\pm0.005$\\

    $\sigma=12.2$             
        & $0.175\pm0.009$
        & $0.054\pm0.005$\\
    \midrule
    $\sigma=4.6$             
        & $0.28\pm0.01$
        & $0.051\pm0.005$\\
    \midrule
    min-$p$  
        & $0.36\pm0.01$
        & $0.089\pm0.006$\\
    prod-$p$ 
        & $\boldsymbol{0.37\pm0.01}$
        & $0.092\pm0.006$\\
    avg-$p$  
        & $\boldsymbol{0.37\pm0.01}$
        & $\boldsymbol{0.130\pm0.007}$\\
    smax-$t$ 
        & $0.28\pm0.01$
        & $0.076\pm0.006$\\    
    \end{tabular}}
    \caption{\textbf{LHCO-6D.} Probability of observing $Z\geq2$ (left column) and $Z\geq3$ (right columns).}
    \label{tab:lhco}
\end{table}
\end{minipage}
\section{Conclusions}\label{sec:5}

In this paper we address the problem of model-selection in ML-based solutions for signal-agnostic searches. By focusing on the NPLM goodness-of-fit test, we show how hyperparameter tuning can introduce biases towards specific signal hypotheses.

We propose to mitigate this effect by performing multiple tests, characterised by different hyperparameters, on the same set of experimental measurements, and combining them into a meta-test in a way that is robust against the look-elsewhere effect. We show that this strategy improves over the baseline proposal in \cite{Letizia:2022xbe}. We observe a more uniform response across multiple signal scenarios, hence an enhanced robustness without loss of sensitivity. In particular, we show that combining individual p-values by selecting the smallest value (the min-$p$ approach) is the most effective method, especially for signals that are hard to detect. This approach involves increased computational requirements as multiple tests have to be performed in place of a single one. However, this cost could be mitigated by an appropriate parallelised strategy.

This work represents a further step towards building unbiased machine learning tools for anomaly detection and hypothesis testing in the context of collider experiments. From this perspective, the strategy proposed in this study goes beyond the NPLM approach and could be tested to combine methods for new physics searches that have been designed to be sensitive to specific families of signals.

In conclusion, our study indicates that the impact of model selection on sensitivity can be leveraged to enhance interpretability, particularly for machine learning models with a limited number of hyperparameters that can be connected to physical priors. This is an interesting direction that we leave for future developments.

\section*{Acknowledgments} 
The authors would like to thank Louis Lyons for his decisive support and pivotal suggestions, and Arthur Gretton and his collaborators for the stimulating conversations. M.L. acknowledges the financial support of the European Research Council (grant SLING 819789).
G.G. acknowledges the financial support of the National Science Foundation under Cooperative Agreement PHY-2019786 (The NSF AI Institute for Artificial Intelligence and Fundamental Interactions, http://iaifi.org/). Computations
in this paper were partially run on the FASRC Cannon cluster supported by the FAS Division of Science
Research Computing Group at Harvard University. 

\bibliographystyle{hunsrt}
\bibliography{references}

\end{document}